\documentstyle[prd,aps,eclepsf]{revtex}
\newcommand{\gtilde}
{~\raisebox{-1ex}{$\stackrel{\textstyle >}{\sim}$}~}

\begin{document}

\title{Effects of Neutrino Oscillation on the Supernova \\ Relic Neutrino
Background}

\author{Tomonori Totani$^1$ and Katsuhiko Sato$^{1,2}$}

\address{
$^1$ Department of Physics, School of Science, The University
of Tokyo, 7-3-1 Hongo, Tokyo 113, Japan. \\
$^2$ Research Center for the Early Universe, School of Science,
The University of Tokyo, 7-3-1 Hongo, Tokyo 113, Japan.}

\maketitle

\vspace{-6cm}
\begin{flushright}
UTAP-237/96 \\
RESCEU-26/96
\end{flushright}

\vspace{5cm}
\begin{abstract}
We investigate to what extent the oscillation or conversion
of neutrinos enhances the expected event rate of the supernova
relic neutrino background (SRN) at the Super-Kamiokande detector (SK).
The SRN $\bar\nu_e$'s can be almost completely exchanged with
$\nu_\mu$-like neutrinos by the MSW oscillation under the inverse 
mass hierarchy with $\Delta m^2 \sim 10^{-8}$--$10^5$ [eV$^2$], 
or by the magnetic moment of Majorana neutrinos with
$\mu_\nu \agt 10^{-12} \mu_B$ and $\Delta m^2
\sim 10^{-4}$--$10^0$ [eV$^2$].
In the standard calculation of the SRN flux, 
the event rate of the SRN $\bar\nu_e$'s at the SK in 
the observable energy range of 15--40 MeV
can be enhanced from 1.2 yr$^{-1}$ to 2.4 yr$^{-1}$
if all $\bar\nu_e$'s are exchanged with $\nu_\mu$-like neutrinos.
The enhancement is prominent
especially in the high energy range (\gtilde 25 MeV). 
In the astrophysically optimistic calculation, the event rate 
becomes as high as 9.4 yr$^{-1}$. Because the theoretical
upper bound of the SRN events without oscillation is about 5 yr$^{-1}$
taking account of the various astrophysical uncertainties,
we might have to resort to the neutrino oscillation if more than
5 events in a year, as well as a significantly harder spectrum,
were observed in the SK. 

\end{abstract}

\vspace{1cm}

There are two astrophysical neutrino sources which have already
been detected until present: the Sun and the Supernova 1987A,
and the source which is expected to become the third 
by the observation of the Super-Kamiokande detector (SK) \cite{SK}
is the supernova relic neutrino background (SRN). The SRN is 
the accumulation of the neutrinos emitted from past supernovae
which have ever occurred in the universe \cite{Guseinov,Bisnovatyi-%
Kogan,Krauss-Glashow-Schramm,Domogatsky,Dar,Woosley-Wilson-Mayle,%
Zhang,Hirata,Totani-Sato1,Totani-Sato-Yoshii}. 
The detection of the SRN
would provide us valuable information on the history of the supernova 
rate and hence the evolution of galaxies, as well as
the properties of the supernova neutrinos. In the earlier paper \cite{%
Totani-Sato-Yoshii}, we calculated the SRN flux and spectrum
by using the standard model of the galaxy evolution and predicted
the realistic event rate of the SRN $\bar\nu_e$'s
at the SK. The expected event rate is 
unfortunately very low: 1.2 yr$^{-1}$ as the central value and 
4.7 yr$^{-1}$ as the optimistic value considering the astrophysical
uncertainties; both are the rate in the observable energy range
of recoil positrons, 15--40 MeV. 
The supernova rate is much higher
in the early phase of the galaxy evolution than in the present, 
but neutrinos emitted from such old supernovae are 
considerably redshifted in energy
and fall below the observable energy range.

However, neutrino oscillation or conversion of the supernova neutrinos
may enhance the expected event rate, and such effects were not 
considered in the previous paper \cite{Totani-Sato1,Totani-Sato-Yoshii}.
We consider here to what extent 
the SRN events can be enhanced by the neutrino oscillation or
conversion. The observable energy window is relatively high compared
with the average energy of supernova neutrinos, 
and the reaction cross section of the detection
($\bar\nu_e p \rightarrow e^+ n$) is proportional to $E_\nu^2$,
where $E_\nu$ is the neutrino energy. Therefore,
if $\bar\nu_e$'s are exchanged with $\nu_\mu$($\bar\nu_\mu$)'s 
or $\nu_\tau$($\bar\nu_\tau$)'s, the SRN event rate
becomes larger because of the higher temperature (or average energy)
of $\nu_\mu$-like neutrinos than $\bar\nu_e$'s \cite{Burrows-etal}.
(In collapse-driven supernovae,
$\nu_\mu$'s, $\nu_\tau$'s and their antiparticles can approximately be
treated as identical particles.) The first of such possibilities is the
vacuum oscillation of $\bar\nu_e \leftrightarrow \bar\nu_\mu$, which
occurs
if the vacuum mixing angle is quite large compared with that of the
quark sector. Although such large mixing is suggested from some of 
the candidates for the solution of the solar neutrino problem,
such as the `just-so' solution \cite{just-so}
or the large-angle MSW solution \cite{large-angle},
the conversion probability of $\bar\nu_e$'s and $\bar\nu_\mu$'s
would at most be 1/2, with the maximum vacuum mixing, after the
phase averaging.
On the other hand, more efficient conversion process exists, that is,
the resonant oscillation or conversion induced by the
effective matter potential in a star
(well known as the MSW effect \cite{MSW}).
However, the ordinary MSW oscillation induced by the mixing of 
the flavor eigenstates and the mass eigenstates is relevant
only to neutrinos and not to antineutrinos, under the direct mass
hierarchy of neutrinos. The most easily detectable flavor,
$\bar\nu_e$, is not exchanged with
$\bar\nu_\mu$ by this ordinary MSW effect.
However, there exist still some possibilities
of almost complete conversion of $\bar\nu_e$'s and $\nu_\mu$-like
neutrinos: one is the MSW effect under the inverse mass hierarchy,
and another is the resonant spin-flavor
conversion of $\bar\nu_e \leftrightarrow \nu_\mu$ induced by the 
flavor changing magnetic moment of Majorana neutrinos.

There are actually some particle-physics models which give
the inverse mass hierarchy \cite{Berezhiani-Chkareuli,%
Petcov-Smirnov,Caldwell-Mohapatra}, and 
some possible phenomenological consequences have also been
discussed\cite{Fuller-Primack-Qian,Raffelt-Silk}.
The MSW resonance condition,
\begin{equation}
\frac{\Delta m^2}{2 E_\nu} \cos 2 \theta =
\sqrt{2} G_F Y_e \frac{\rho}{m_p} \ ,
\label{res-cond}
\end{equation}
can be satisfied with $\Delta m^2 \alt 10^5$ [eV$^2$]
above the neutrino sphere in supernovae, where
$\Delta m^2 = m^2_{\bar\nu_e} - m^2_{\bar\nu_\mu}$ (defined as 
positive under the inverse mass hierarchy), $\theta$ is the
vacuum mixing angle, $Y_e$ the electron number fraction per nucleon,
$m_p$ the proton mass,
and $\rho$ the density. We use here $\rho \sim 10^{11}$ [g/cm$^3$] and
$Y_e \sim 0.37$ at the neutrino sphere\cite{Fuller-etal},
and also use $E_\nu \sim$ 20 MeV and $\cos 2 \theta \sim 1$.
The another condition necessary for the conversion, i.e.,
the adiabaticity condtion is given by
\begin{equation}
\frac{\Delta m^2 \sin^2 2 \theta}{2 E_\nu \cos 2 \theta}
\agt \frac{1}{n_e} \left| \frac{dn_e}{dr} \right| \sim 
\frac{1}{\rho} \left| \frac{d\rho}{dr} \right|
\ ,
\label{ad-cond}
\end{equation}
at the resonance layer, where $r$ is the radius from the center of the 
star, and $n_e$ the number density of electrons. 
This condition is satisfied around the neutrino sphere
when $\sin^2 2 \theta \agt 3 \times 10^{-9}$, where we assume
the right-hand side of Eq. (\ref{ad-cond}) is $\sim$ (30 km)$^{-1}$,
i.e., the typical radius of the neutrino sphere.
When $\Delta m^2$ is smaller than $\sim 10^5$ [eV$^2$],
the resonance occurs in more outer region in the star,
and necessary mixing, $\sin^2 2 \theta$ for the adiabatic condition
becomes larger.
In order to show this, we give here a rough estimate of the adiabaticity.
Let us assume the density profile is described by a power-law:
$\rho \propto r^{-k}$. If we neglect the change in $Y_e$, 
the resonance layer, $r_{res}$, scales as $r_{res}^{-k} \propto 
\Delta m^2$, from Eq. (\ref{res-cond}). On the other hand,
the right-hand side of the adiabaticity condition, Eq. (\ref{ad-cond}),
scales as $\propto r^{-1}$. Therefore, from the relation of $r_{res}$
and $\Delta m^2$, the right-hand side of Eq. (\ref{ad-cond}) scales
as $\propto (\Delta m^2)^{1/k}$.
Then it is easy to see that the necessary mixing, $\sin^2 2 \theta$
scales as $\propto (\Delta m^2)^{1/k-1}$, i.e.,
\begin{equation}
\sin^2 2 \theta \agt 3 \times 10^{-9} \left( \frac{\Delta m^2}
{10^5 [{\rm eV^2}]} \right)^{\frac{1}{k}-1} \ .
\label{ad-cond2}
\end{equation}
The numerical calculation\cite{Woosley-Weaver} shows that the 
value of $k$ lies in the range of 2--3, and hence the lower bound for
$\Delta m^2$ to keep the right-hand side of Eq. (\ref{ad-cond2})
below $\sim$ 1 is $\Delta m^2 \agt 2 \times 10^{-8}$ [eV$^2$]
($k=3$) or $10^{-12}$ [eV$^2$] ($k=2$). Therefore the mass range
relevant to the MSW oscillation in supernovae is 
$\Delta m^2 \sim 10^{-8}$--$10^{5}$ [eV$^2$],
and with these values of the parameters, the conversion of 
$\bar\nu_e$'s and $\bar\nu_\mu$'s in the SRN spectrum is possible.

\begin{figure}[tpb]
\begin{center}
\epsfile{file=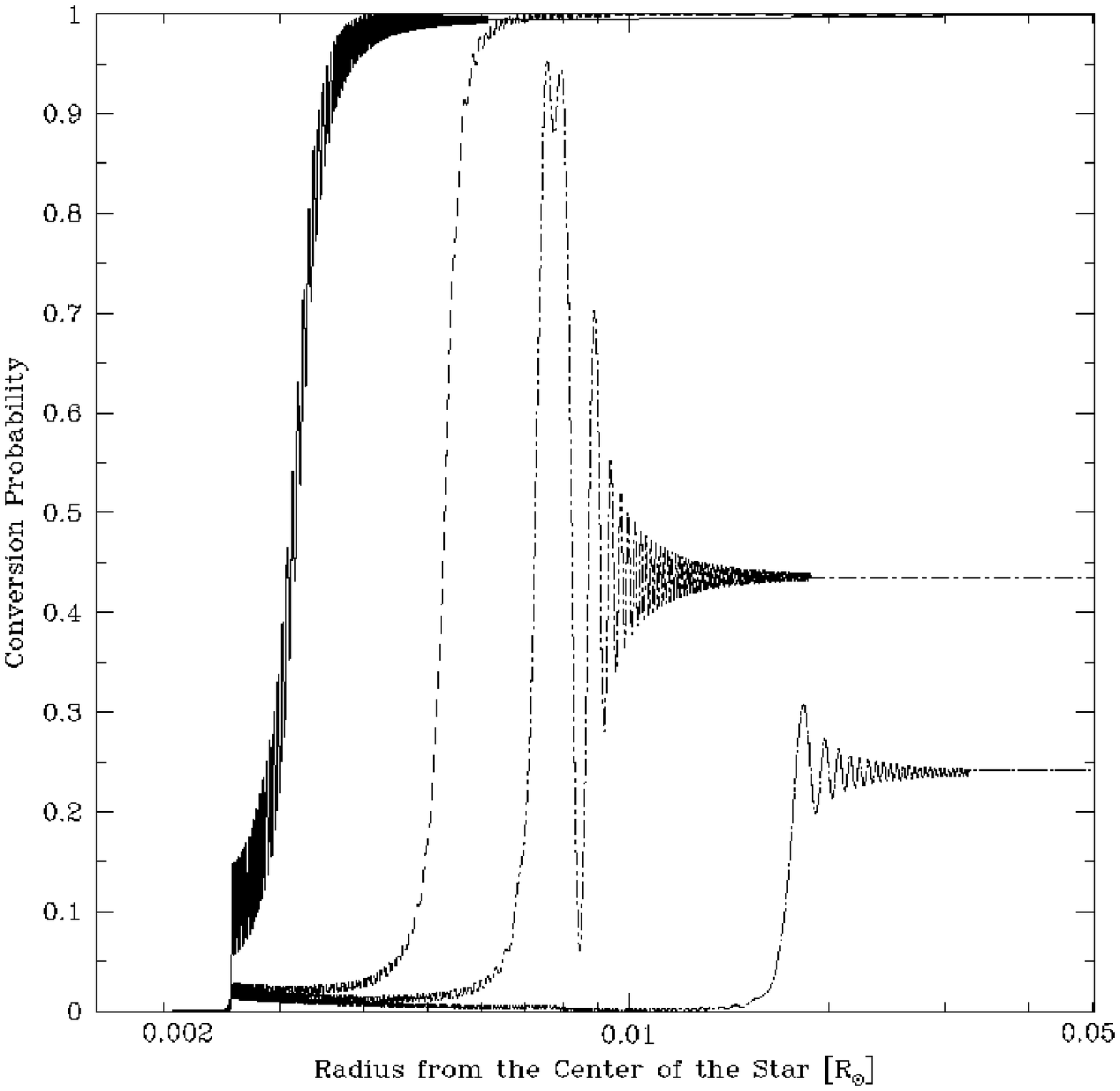,width=12cm}
\end{center}
\caption{
The evolution of the conversion probability of $\bar\nu_e
\leftrightarrow \nu_\mu$ in the isotopically neutral region of
a massive star, due to the magnetic moment of Majorana
neutrinos. The 4 $M_\odot$ helium core model of Nomoto and 
Hashimoto\protect\cite{Nomoto-Hashimoto} is used, and $\mu_\nu = 10^{-12}
\mu_B$ is assumed. The strength of magnetic fields is assumed to
be $B_0 = 5 \times 10^9$ [Gauss], where $B_0$ is $B$ at the surface
of the iron core, and a global magnetic dipole is also assumed.
The values of $\Delta m^2 / E_\nu$ are $3\times 10^{-4}$ (solid line),
$1\times 10^{-4}$ (dashed line), $5\times 10^{-5}$ (dot-short-dashed
line), and $1\times 10^{-5}$ (dot-long-dashed line),
in units of [eV$^2$/MeV].}
\label{fig:sf-osc}
\end{figure}

Also for the magnetic moment of neutrinos, there are
some practical particle-physics 
models\cite{Fukugita-Yanagida,Babu-Mathur} which give the magnetic 
moment of about $\mu_\nu \sim 10^{-12} \mu_B$, and the phenomenological 
consequences of the resonant conversion in the Sun or supernovae have
been discussed\cite{Lim-Marciano,Akhmedov,Voloshin,Akhmedov-Berezhiani,%
Peltoniemi,Athar-Peltoniemi-Smirnov,Totani-Sato2}.
The conversion of $\bar\nu_e$'s and $\nu_\mu$'s
(or $\nu_\tau$'s) is especially effective in the isotopically 
neutral region of massive stars, i.e., above the iron core and 
below the hydrogen 
envelope\cite{Athar-Peltoniemi-Smirnov}.
For supernovae with the solar metallicity,
the relevant mass range\cite{Totani-Sato2} is $\Delta m^2 \sim 
10^{-4}$--$10^{0}$ [eV$^2$], and necessary magnetic interaction is 
$\mu_{\bar\nu_e\nu_\mu} \agt 10^{-12}
(10^9{\rm G}/B_0) \mu_B$, where $B_0$ is the magnetic field
at the surface of the iron core and $10^9$ [Gauss]
is an reasonable value inferred from the observed magnetic fields
of white dwarfs\cite{Chanmugam}. (Conversion may occur also with 
$\Delta m^2$ larger than $10^0$ [eV$^2$], but this corresponds to
the resonance in the collapsed iron core, and the detailed calculation
is quite difficult.)
The metallicity of the progenitor affects the conversion 
probability\cite{Totani-Sato2}, and the conversion can occur also with
smaller $\Delta m^2$ than the above range in lower metallicity stars.
Such effects may be important in supernovae in the early phase of 
galaxy evolution.
However, it should be noted that the magnetic moment of neutrinos
interacts only with the transverse magnetic fields.
Considering the global structure of magnetic fields in the collapsing 
star, the conversion probability changes with the direction of 
neutrinos. The conversion probability after averaging over direction,
however, can be close to the unity, with some appropriate parameters.
In Fig. \ref{fig:sf-osc}, the examples of the evolution of 
the conversion probability due to this mechanism are shown,
by using the precollapse model of Nomoto and Hashimoto 
(1988)\cite{Nomoto-Hashimoto}.

Considering these possibilities, we calculate
the SRN flux and event rate at the SK under the condition of the 
complete conversion, i.e., $P=1$, where $P$ is the conversion 
probability of $\bar\nu_e$'s and $\nu_\mu$-like neutrinos.
Because the SRN is the linear superposition of the (redshifted)
spectrum of supernova neutrinos, it is easy to get the 
SRN spectrum with any values of $P$:
\begin{equation}
F(E_\nu) = (1-P) F_{\bar\nu_e}(E_\nu) + P F_{\nu_\mu}(E_\nu) \ ,
\end{equation}
where $F$ is the differential flux of the SRN and $E_\nu$
is the neutrino energy. We assume that $P$ is independent of 
$E_\nu$ throughout this paper. In fact, in the both mechanisms
described above, the conversion probability generally depends
on $E_\nu$. Because only the neutrinos with high energy
($\agt 15$ MeV) can be observed by the SK, the conversion
in the whole energy range is not necessary to enhance the event rate
of the SK, but the conversion above 15 MeV is sufficient.

It should also be noted that strong conversion of $\bar\nu_e$'s and 
$\nu_\nu$-like neutrinos is disfavored by the data
of neutrinos from SN1987A, observed in Kamiokande II
\cite{Kam} and IMB \cite{IMB}. Smirnov, Spergel, and
Bahcall \cite{Smirnov-Spergel-Bahcall}
concluded that $P > 0.35$ is excluded by the 99 \% confidence level.
Although Kernan and Krauss \cite{Kernan-Krauss} arrived at the 
opposite conclusion
that a significant oscillation is favored by the data, the
obtained parameters of supernova neutrino spectrum are quite
different from the theoretically plausible parameters. 
Jegerlehner, Neubig, and Raffelt \cite{Jegerlehner-Neubig-Raffelt} 
also concluded by using the maximum likelihood method
that the fitted temperature of SN1987A data should be less than
4.7 MeV with the confidence level of 95.4 \%, which is considerably
small compared with the theoretically predicted $\nu_\mu$
temperatures. However, because of the statistical uncertainties,
one cannot completely exclude the possibility of the full conversion.
Actually, the above values of the confidence level are determined by
the Kolmogolov-Smirnov test in Ref.\cite{Smirnov-Spergel-Bahcall}
and by the difference of the value of the likelihood function in 
Ref.\cite{Jegerlehner-Neubig-Raffelt}, both of the two methods
suffer considerable uncertainty when the sample size is small.
Therefore, we consider the case of the complete conversion 
of $\bar\nu_e$'s and $\nu_\mu$-like neutrinos, as the simplest model.

\begin{figure}
\begin{center}
\epsfile{file=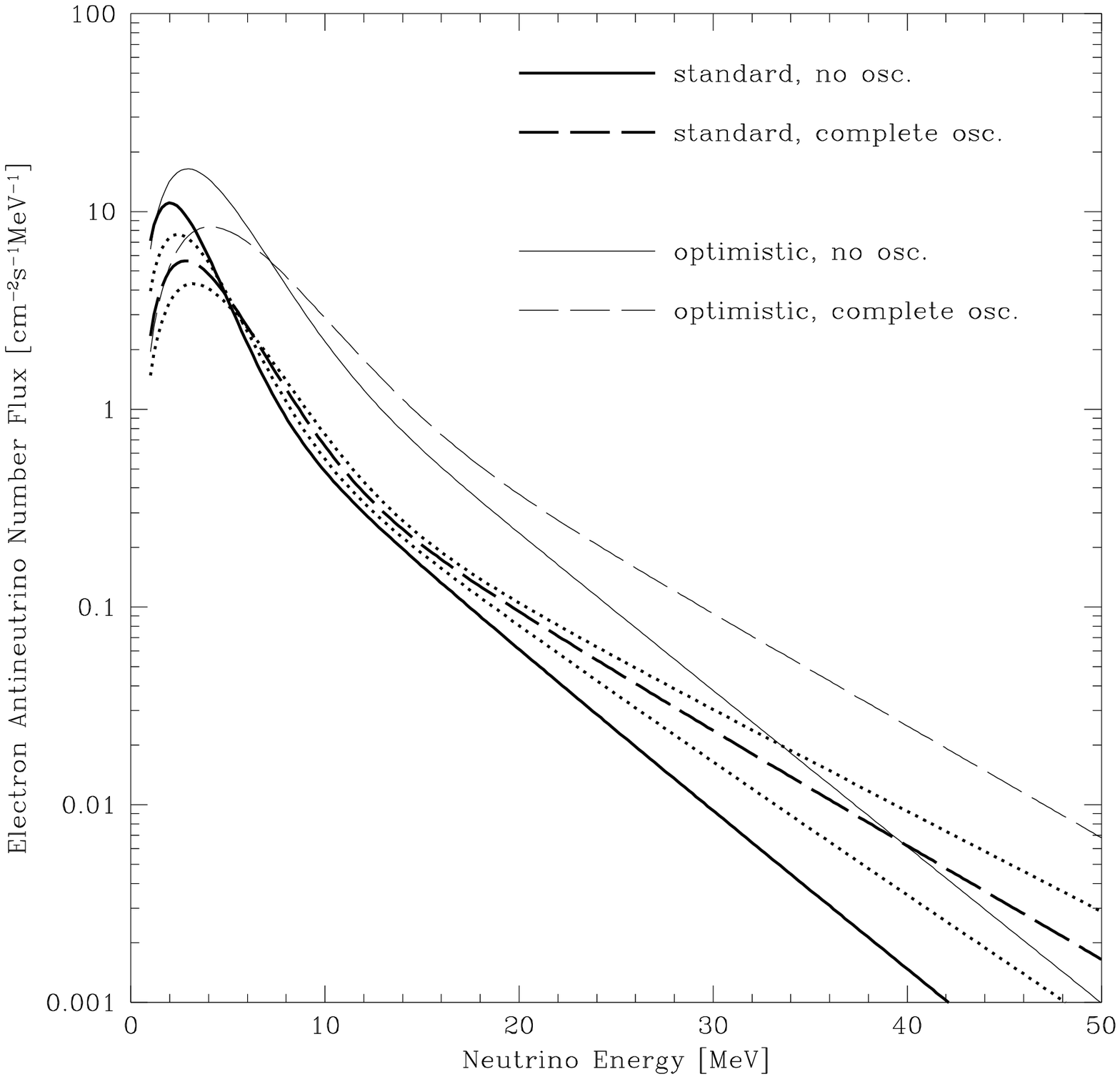,width=12cm}
\end{center}
\caption{
The differential number flux of the SRN as a function
of neutrino energy. The thick-solid line shows the standard 
spectrum without any oscillation or conversion of neutrinos.
The thick-dashed line is the SRN spectrum when the original
$\bar\nu_e$'s are completely exchanged with $\nu_\mu$-like neutrinos.
The ratio of temperature, $T_{\nu_\mu}/T_{\bar\nu_e}$ is assumed to be
1.4, and the thick-dotted lines are the spectra with 
$T_{\nu_\mu}/T_{\bar\nu_e} = 1.6$ (upper) and 1.2 (lower).
The thin-solid line is the optimistic SRN flux considering the 
astrophysical uncertainties (no conversion of neutrinos). 
The thin-dashed line is 
the SRN flux using the optimistic model, and with the complete 
conversion of $\bar\nu_e$'s and $\nu_\mu$-like neutrinos, again
assuming $T_{\nu_\mu}/T_{\bar\nu_e}$ = 1.4.}
\label{fig:srn-flux}
\end{figure}

In the earlier calculation which did not take account of
the possibility of neutrino oscillation \cite{Totani-Sato-Yoshii},
the spectrum of the neutrino emission from a supernova is
assumed to obey the Fermi-Dirac distribution with zero chemical
potential \cite{Burrows-etal}. All stars in the mass range
of 8--60 $M_\odot$ are assumed to end their life with a collapse-driven
supernova, and the region is divided into the three sub-regions:
8--12.5, 12.5--20, and 20--60 $M_\odot$. The total energy 
($E_{\bar\nu_e}$)
and temperature ($T_{\bar\nu_e}$) of $\bar\nu_e$'s emitted
from a supernova in 
these three mass regions are represented by those of the supernovae 
with the mass of 10, 15, and 25 $M_\odot$, respectively, 
by using the results of Woosley, Wilson,
and Mayle \cite{Woosley-Wilson-Mayle}. Their calculation gives
($E_{\bar\nu_e}$, $T_{\bar\nu_e}$) = (4.8, 4.0), (6.0, 5.0), and
(11, 5.3) in units of ($10^{52}$ ergs, MeV), for 10, 15, and 25
$M_\odot$ stars, respectively. In addition to this modeling,
here we assume that the total energy of neutrinos is the
same for $\bar\nu_e$'s and $\nu_\mu$-like neutrinos while the 
temperature of $\nu_\mu$-like neutrinos is higher than $T_{\bar\nu_e}$
by a factor of 1.4 for all supernovae, in order to include the 
effect of neutrino oscillation.
This simple model is inferred from the simulation of neutrino emission
from a supernova of about 20 $M_\odot$
up to 18 [sec] after the collapse by Wilson and Mayle \cite{Wilson}
(see also Ref.\cite{Woosley-etal}). Following their calculation,
the total energy and the average energy of each neutrino
are (4.6, 15.3) for $\bar\nu_e$'s
and (4.9, 21.7) for $\nu_\mu$-like neutrinos, 
again in units of ($10^{52}$ ergs, MeV). With these assumptions,
we can calculate the SRN flux and spectrum in the case that 
all or a part of $\bar\nu_e$'s are exchanged with
$\nu_\mu$-like neutrinos. Fig. \ref{fig:srn-flux} shows the calculated SRN
differential number flux. The thick-solid line is the standard 
SRN spectrum obtained in the previous work \cite{Totani-Sato-Yoshii}
without any neutrino oscillation or conversion. The thick-dashed
line is also the standard SRN spectrum, but when all $\bar\nu_e$'s 
are completely exchanged with $\nu_\mu$-like neutrinos ($P=1$). 
The thick-dotted lines
are the same with the thick-dashed line, but with different values
of $T_{\nu_\mu} / T_{\bar\nu_e}$ ratio: 1.2 (lower line) and 1.6
(upper line). By these lines, one can estimate how the uncertainty
of $T_{\nu_\mu} / T_{\bar\nu_e}$ ratio affects the SRN spectrum.
The thin-solid line is the `optimistic' SRN spectrum (no conversion), 
with a larger value of the Hubble constant and luminosity
density of galaxies, and also using a different model of galaxy 
evolution which enhances the SRN flux in high energy region (see
Ref.\cite{Totani-Sato-Yoshii}). The thin-dashed line is the same
with the thin-solid line, but with the condition of $P=1$.
As expected, the neutrino oscillation increases the neutrino 
energy and the spectrum becomes harder. This hardening is expected 
to enhance the event rate at the SK detector. 

\begin{figure}
\begin{center}
\epsfile{file=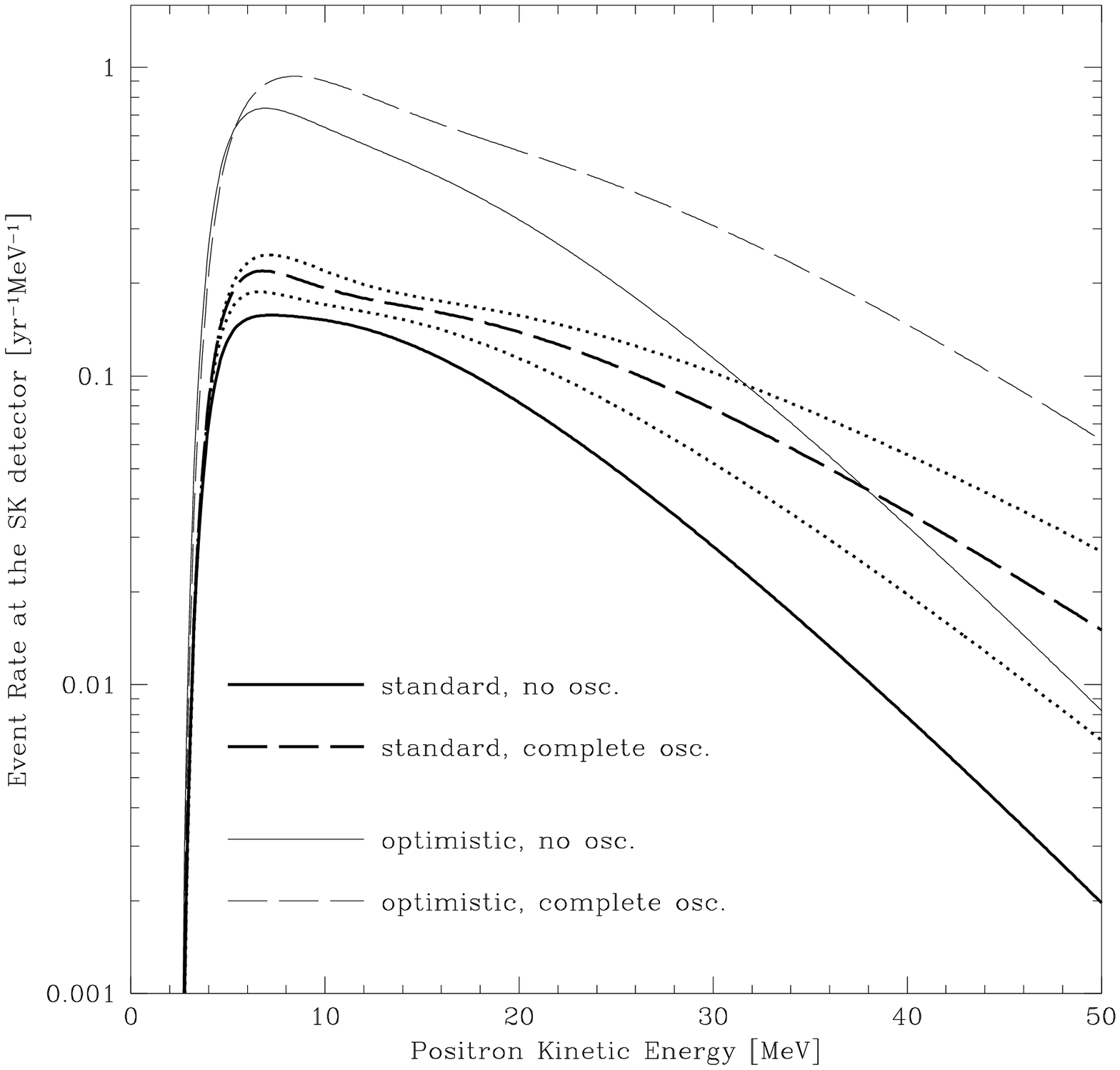,width=12cm}
\end{center}
\caption{
The expected event rate of the SRN $\bar\nu_e$'s
at the Super-Kamiokande detector, as a function of the kinetic energy
of recoil positrons.  All the lines correspond
to those in Fig. \protect\ref{fig:srn-flux}, in which the SRN flux is shown.
Note that the observable energy range is about 15--40 MeV
because of the other background events.}
\label{fig:srn-evrt}
\end{figure}

Fig. \ref{fig:srn-evrt} shows the expected event rate at the 
SK as a function of kinetic energy of recoil positrons, which
are produced by the reaction $\bar\nu_e p \rightarrow n e^+$.
All the lines correspond to those of the SRN flux shown in
Fig. \ref{fig:srn-flux}. We use 9.72 $\times 10^{-44} E_e p_e$
cm$^2$ as the cross section of the $\bar\nu_e$
absorption reaction, where $E_e$ and $p_e$ are the energy and 
momentum of recoil positrons, measured in MeV. 
The fiducial volume of the SK for
the SRN observation is 22,000 tons (same with the solar neutrino
observation), and the detection efficiency 
is also taken into account, which is provided by 
Totsuka \cite{Totsuka-private}. It is apparent from this figure
that the event rate is enhanced by the neutrino oscillation, 
especially in the high energy region (\gtilde 25 MeV).
In Table \ref{table:sk-events}, the energy-integrated event rate
in the observable energy range of 15--40 MeV is shown for 
the models calculated in this paper. With $T_{\nu_\mu} /
T_{\bar\nu_e} = 1.4$, the event rate of the standard
calculation is enhanced up to 2.4 yr$^{-1}$ from 1.2, and the optimistic
calculation up to 9.4 yr$^{-1}$ from 4.7, if the conversion is complete. 
The important property of 
the effect of oscillation is that the enhancement of the event rate
is prominent in the higher energy range. Therefore, the event rate
in the range of 25--40 MeV is also shown in the table. 
About four events in a year are possible in this high energy range
if we use the optimistic model. 

In conclusion, the optimistic event rate of 4.7 yr$^{-1}$ can be
raised up to 9.4 yr$^{-1}$ by the {\it complete}
neutrino oscillation or conversion
of $\bar\nu_e$'s and $\nu_\mu$-like neutrinos. Because the rate 4.7
yr$^{-1}$ in 15--40 MeV is considered to be a theoretical upper bound 
taking account of various astrophysical uncertainties, we might have to
resort to the neutrino oscillation or conversion 
if more than 5 events were observed
in the SK detector in a year. In this case, the events in the high energy
range (\gtilde 25 MeV) will be especially enhanced compared with the 
event rate spectrum without any oscillation or conversion.
The theoretical 
upper bound on the SRN events in the SK becomes 9.4 yr$^{-1}$ when
the neutrino oscillation is taken into account, but since the 
complete conversion seems rather unlikely considering the 
SN1987A data, this upper bound should be considered to be quite
conservative one.

The authors would like to thank J. R. Wilson for providing us
their numerical results of collapse-driven supernova simulation,
and useful comments.  They would also like to thank S. E. Woosley
and M. Hashimoto, for providing us the precollapse models of
massive stars.
They are also grateful to Y. Totsuka for the information on 
the detection efficiency of the Super-Kamiokande. They are
also indebted to H. Nunokawa for the careful reading of this 
manuscript. This work has been
supported in part by the Grant-in-Aid for COE Research (07CE2002) and
for Scientific Research Fund (05243103, 07640386 and 3730) of the Ministry
of Education, Science, and Culture in Japan.

\vspace{3cm}

\begin{table}
  \caption{Expected Event Rate at the Super-Kamiokande Detector}
  \label{table:sk-events}
  \begin{center}
  \begin{tabular}{cccc}
    && \multicolumn{2}{c}{Event Rate [yr$^{-1}$]} \\
    \cline{3-4}
    Astrophysical Model & $T_{\nu_\mu} / T_{\bar\nu_e}$ & 
                                   15--40 MeV & 25--40 MeV \\
    \hline
    Standard   & 1    & 1.2 & 0.35 \\
               & 1.2  & 1.8 & 0.66 \\
               & 1.4  & 2.4 & 1.0  \\
               & 1.6  & 2.9 & 1.35 \\
    Optimistic & 1    & 4.7 & 1.42 \\
               & 1.4  & 9.4 & 4.0 \\
  \end{tabular}
  \end{center}
Note.--- The conversion probability, $P$, is assumed to be the unity, i.e.,
the complete conversion of $\bar\nu_e$'s and $\nu_\mu$-like neutrinos.
The energy ranges are those of the kinetic energy of recoil positrons.
\end{table}

\end{document}